\documentstyle[12pt]{article}
\begin{document}

\renewcommand{\Large}{\normalsize}
\renewcommand{\huge}{\normalsize}
\def\be{\begin{equation}}
\def\en{\end{equation}}
\def\bear{\begin{eqnarray}}
\def\enar{\end{eqnarray}}

\begin{titlepage}
\baselineskip .3in

\begin{center}
{\LARGE{\bf Can Gravitational Waves Prevent Inflation?} }

\vskip .5in

{\sc  Hisa-aki Shinkai}$^1$ and {\sc Kei-ichi Maeda}$^2$\\[1em]
 {\em Department of Physics, Waseda University,
Shinjuku-ku, Tokyo 169-50, Japan}
{}~\\
{}~\\

May 18th, 1993

\end{center}
\vfill
\begin{abstract}
\baselineskip .35in

To investigate the cosmic no hair conjecture, we
 analyze numerically 1-dimensional plane symmetrical inhomogeneities
due to
gravitational waves in  vacuum spacetimes with a positive cosmological
constant.  Assuming periodic gravitational
pulse waves initially, we  study the time evolution  of those waves
and the nature of their collisions.
As  measures of inhomogeneity  on each hypersurface, we use the
3-dimensional Riemann invariant ${\cal I}\equiv
{}~^{(3)\!}R_{ijkl}~^{(3)\!}R^{ijkl}$ and the electric and magnetic
parts of
the Weyl tensor.  We find a temporal growth of the curvature in the
waves'
collision region, but the overall  expansion of  the universe later
overcomes this effect. No singularity appears and the
 result is  a  ``no hair" de Sitter spacetime.
The waves we study have amplitudes between $0.020\Lambda \leq
{\cal I}^{1/2} \leq 125.0\Lambda$ and widths between $0.080l_H
\leq l \leq 2.5l_H$,  where  $l_H=(\Lambda/3)^{-1/2}$, the horizon
scale of de
Sitter spacetime. This supports the cosmic no hair conjecture.\\

 \end{abstract}

\vfill

1~~electronic mail: 62L508@jpnwas00.bitnet \\

2~~electronic mail: maeda@jpnwas00.bitnet ~or~ maeda@cfi.waseda.ac.jp
\\
\end{titlepage}

\vspace{.5 cm}
\normalsize
\baselineskip = 23pt

The widely accepted  inflationary universe model\cite{inflation} is an
attempt to solve many long-standing cosmological difficulties---such
as the flatness,  horizon,  and monopole problems---in the standard
big-bang model.   The explanation of the present isotropy and
homogeneity of the universe are one of those problems.
The inflationary scenario  may resolve this problem by introducing a
rapid cosmological expansion. But the scenario is usually studied in
an isotropic and homogeneous spacetime, so that   to determine whether
inflation really isotropizes and homogenizes the universe and to see
whether  inflation is  generic,  we have to study an inflationary
model in anisotropic and inhomogeneous spacetimes as well.

To explain the generality of inflation, the  so-called cosmic
no hair conjecture has proposed\cite{nohair}.  The conjecture is:
``All initially expanding universes with  positive cosmological
constant $\Lambda$ approach the de Sitter spacetime asymptotically".
There exist, however, simple counter examples.  For example, a closed
Friedmann-Robertson-Walker spacetime with a positive $\Lambda$
recollapses if the initial energy density is large enough. On the
other hand, the de Sitter spacetime is stable against perturbations
and  many other models also  support this conjecture, so that
 we expect we can  prove a suitably refined version of this
conjecture.

 For homogeneous but anisotropic spacetimes, Wald\cite{wald}
 showed that, except in the Bianchi IX case, initial expansion and
positive $\Lambda$ force an approach to the  de Sitter spacetime
within one
Hubble expansion time $\tau_H = \sqrt{\Lambda / 3}$. And for Bianchi
IX, the
additional initial condition that $\Lambda$ be larger than one third
of the
maximum value of 3-dimensional Ricci scalar leads to the
same conclusion\cite{wald,kitamae}.
 For inhomogeneous spacetimes, on the other hand, the only practical
method we have at present is to solve the Einstein equations
numerically. Assuming a spherically symmetric spacetime, Goldwirth and
Piran\cite{GP}  studied the  behavior of inhomogeneous distributions
of a scalar field.  They found a sufficient criterion for inflation
that
the physical scale of inhomogeneity of the scalar field  be
larger than  several times the horizon scale.

We may, however,  wonder whether this result is valid for cases with
other
symmetries or for more generic spacetimes.  Assuming a plane
symmetric spacetime, Kurki-Suonio {\it et. al.}\cite{suonio} studied
new
inflationary model and found that inflation occurs only when the
potential is
flat enough.  The spacetime they studied is, however, not sufficiently
inhomogeneous in order to investigate the ``no hair conjecture".
This is the first reason why we study a plane symmetric spacetime in
this paper.

Besides the inflaton field, there exist  other sources of
inhomogeneity, one of which is gravitational waves.
Although any linear gravitational waves in de Sitter space will decay
exponentially and desappear soon\cite{weakgw},
we do not know what will happen if strong gravitational waves exist.
If the strong gravitational waves are localized, those may collapse
into
black holes.  Even if those are not localized as a plane symmetric
case, such
a strong gravitational field may evolve into a naked singularity.
In fact,  Szekeres and
Khan-Penrose\cite{colwave} showed that a collision of two plane
symmetric
gravitational waves in Minkowski background spacetime will form a
singularity. We may wonder what will happen with strong gravitational
waves in de Sitter
background.  A singularity may be formed and prevent inflation.   In
this report, we study
such an inhomogeneity due to gravitational waves.   Because they
cannot exist in a spherically symmetric spacetime, gravitational wave
inhomogeneities may play a different role in the homogenization
process
 from that of the inflaton field analyzed previously. This is the
second reason why we study a plane symmetric spacetime here.

Recently, Nakao {\it et al.} examined time symmetrical initial
data for Brill waves in an axial symmetrical spacetime with
cosmological constant and found that waves with large
gravitational mass does not always provide trapped surfaces\
cite{nakaogw}.
They also found that a dust sphere with
large gravitational mass in a background de Sitter spacetime
does not collapse to form a black hole spacetime\cite{nakaodust}, and
that there exists an upper bound on the area of the apparent horizon
of a black hole  in an asymptotically de Sitter
spacetime\cite{shiromizu}, so that we may conjecture that a large
inhomogeneity
does not necessarily  prevent cosmological constant-driven inflation.
{}From these results, it seems plausible that in an inflationary era,
inhomogeneities will simply evolve into many small black holes in a
background de
Sitter universe; the only worrisome possibility is that a naked
singularity
might form.  Disposing of this worrisome possibility  may thus
become the main problem in the study of the cosmic no hair conjecture.
Our
1-dimensional case seems well suited to address this problem.
This is the third reason since a singularity formed in a plane
symmetric
spacetime is always naked.

For the above three reasons,  we have carried
out a numerical study of  inhomogeneities due to plane-symmetric
gravitational
waves with a cosmological constant.   Since we  treat a plane
symmetric
spacetime and gravitational waves, our simulation may be thought of a
complementary work to the spherically symmetric one discussed by
Goldwirth and
Piran.   Since we have not studied an inflaton field in this paper,
however, our result may not
be applied to a realistic universe as it is. (See also our comment in
the references).

To analyze the behavior of 1-dimensional inhomogeneities due to
gravitational
waves, we assume spacetime is plane symmetric and vacuum. We also
assume that
a  positive cosmological constant $\Lambda$ exists.
Under these assumptions, we examine whether  such a spacetime leads to
an
 inflationary era, and whether such  initial inhomogeneities and
anisotropies
 smooth out during inflation periods.

We use the ADM formalism  to solve the Einstein equation
\be
R_{\mu \nu}-{1 \over 2} g_{\mu \nu} R+\Lambda g_{\mu \nu} =0, \
label{einstein}
\en
with the metric
\be
ds^2 = -(\alpha^2 -\beta^2/\gamma_{xx})dt^2+2\beta dtdx
+\gamma_{xx}dx^2+\gamma_{yy}dy^2+\gamma_{yz}dydz+\gamma_{zz}dz^2,
\label{metric0}  \label{metric}
\en
where the lapse function
$\alpha$, the shift vector $\beta_i=(\beta,0,0)$, and the 3-metric
$\gamma_{ij}$  depend  only on the
time $t$ and propagation direction coordinate $x$.

We use the Hubble expansion time $\tau_H=\sqrt{\Lambda / 3}$ as our
time
unit, which is a characteristic expansion time of the expanding
Universe.  And our
unit of length is also normalized to the horizon length of the de
Sitter universe
$l_H=1/\sqrt{\Lambda/3}$.

Our simulation procedure follows Nakamura {\it et
al.}\cite{adm3}:
(i) Determine initial values by solving two constraint
equations.
(ii) Evolve time slices by
using the  dynamical equations.
(iii) Check the results of (i) and (ii)  using the
two constraint equations on every time slice.

To determine the initial data, we use the 5th order Runge-Kutta method
(Fehlberg
method), and to integrate  dynamical equations, we use a finite
difference method
with 400 meshes. In our calculation, the maximum
errors in the Hamiltonian and momentum constraint equations are
$O(10^{-4})$
and $O(10^{-6})$ on the initial hypersurface,  respectively,
and these accuracies are maintained even after  evolution.

 We use the  York-O'Murchadha's conformal approach\cite{initial0} to
get
initial values. Setting
$\gamma_{ij}= \phi^4\hat{\gamma}_{ij}$, the
Hamiltonian constraint equation
becomes
\be
{}~8~^{(3)\!}\hat{\Delta}
\phi=~^{(3)\!}\hat{R}\phi-\hat{A}_{ij}\hat{A}^{ij}\phi^{-7} +2({1 \over
3}
\hat{K}^2 - \Lambda)\phi^5 \label{shoki1}
\en
where $^{(3)\!}\hat{\Delta}$ is
the 3-dimensional Laplacian of a conformal metric $\hat{\gamma}_{ij}$,
$~^{(3)\!}\hat{R}$ is the 3-dimensional scalar curvature, $\
hat{A}_{ij}$ is the
trace-free part of the extrinsic curvature $\hat{K}_{ij}$ and $\
hat{K}(=K)$ is a
trace part of $\hat{K}_{ij}$. The quantities with a hat denote
physical variables in
the conformal frame.
 In this approach, $\hat{\gamma}_{ij}, \hat{K}$ and the transverse-
traceless(TT)
part of the extrinsic curvature $\hat{K}^{TT}_{ij}$ are  left to our
choice.

 We assume  constant mean curvature on the initial
hypersurface,
\be
 \hat{K}=-\sqrt{3 \Lambda}(1+\delta_K)= {\rm const.}~~~~~{\rm on}~\
Sigma(t=0),
\label{tracecon} \en
where $\delta_K$ is a constant. Here, $\delta_K$ is introduced to find
the
periodic gravitational waves.  $\delta_K=0$ is inconsistent with the
presence of
gravitational waves in the initial data.
The reason is, when we have
pulse-like inhomogeneities under  periodic boundary condition, there
are two
energy sources; one is a positive cosmological constant and the other
is the
energy of pulse waves.  These energy give contributions to the
expansion of the
universe, {\it i.e.} the first term in the r.h.s. of (\ref{tracecon})
is due
to the cosmological constant, while the second one ($\delta_K$) is due
to the
pulse waves.

 With this slicing, the momentum constraint equation
becomes trivial and $\hat{A}_{ij}=\hat{K}^{TT}_{ij}$.
 If we also impose  $\phi \rightarrow 1$
at the numerical boundary edges, the value of $\delta_K$ is fixed; it
is
about $O(10^{-2})$ in the physical frame.

For coordinate conditions, we impose  a geodesic slicing
condition such as $\alpha =1$  and   $\beta=0$. This slicing condition
entails the
risk that our numerical hypersurface may hit a  singularity and
 stop there. If
 no singularity appears, however,  then it may be the best coordinate
condition  for revealing whether  de Sitter space emerges as a result
of time evolution.

As for the initial data, we consider
 two complementary cases:

{\bf [case 1]} The inhomogeneities reside in the 3-metric $\hat{\
gamma}_{ij}$
and  $\hat{K}_{ij}^{TT} =0$,

{\bf [case 2]} The 3-metric is conformally flat and the
inhomogeneities
reside in  $\hat{K}_{ij}^{TT}$.

\noindent
In this Brief Reports,  we present mainly the results of simulations
for the [case 1],
and  we  examine
 a pulse-like wave propagating  in the $x$-direction, and  expressed
by the
metric in York's conformal frame as
\be
{\mbox{\rm diag }}(\hat{\gamma}_{ij})=(1, 1+a~{\rm e}^{-(x/x_0)^2},
1),
  \label{try1}
\en
where $a$ and $x_0$ are free parameters.  Varying these parameters,
which we interpret
as the wave's amplitude and width,  we study the effects of
a family of gravitational pulse waves on the spacetime structure.

As a measure of inhomogeneity, we first use the Riemann invariant
scalar
${\cal I} \equiv ~^{(3)\!}R_{ijkl}~^{(3)\!}R^{ijkl}$, where
$~^{(3)\!}R_{ijkl}$ is the Riemann tensor of the 3-metric on $\Sigma$,
and we
introduce a dimensionless variable normalized by the  cosmological
constant,
 \be
{\cal C}(t,x)\equiv {{\cal I}^{1/2} \over
\Lambda}~~~~~{\rm on} ~\Sigma(t), \label{curvature}
\en
and call it  the  ``curvature" hereafter.  We estimate the magnitude
of
the inhomogeneities in the 3-space $\Sigma$ by the
maximum value of this ``curvature"  on each slice, {\it i.e.} ${\cal
C}_{\rm
max}(t) = \max\{ {\cal C}(t,x)~|_{ ~x \in \Sigma }\} $.

As we mentioned a moment ago, a pulse-like wave has two characteristic
physical dimensions, a width and an amplitude. We use ${\cal C}_{\rm
max}(t)$ as an
amplitude measure, and we define the width $l(t)$
by the proper distance between two points where the trace $\gamma$ of
3-metric $\gamma_{ij}$ (the square of the 3-volume) decreases by  half
from its
maximum value  $\gamma_{max}$.
When $a$ and $x_0$ are given, the initial form of a gravitational
wave is determined, and the width $l(0)$ and ${\cal C}_{\rm max}(0)$
may be calculated.

 In Figure 1, we show a typical
 time evolution of this model for the case  $a=-0.10,
x_0=0.20;  l=0.33,{\cal C}_{\rm max}(0)=1.25$.  We see
that  the wave given on the initial surface propagates
both in the $+ x$ and $- x$ directions, and the
``curvature" seems to be superposed in the interacting
region of the waves. The spacetime, however, finally
succumbs to the overall expansion driven by  the
cosmological constant, and becomes virtually
indistinguishable from de Sitter spacetime  within  one
Hubble expansion time  $\tau_H$.

We also use  4-dimensional variables as a measure to see the
homogenization
process.  In the vacuum spacetime, if the  Weyl tensor $C_{\mu \nu \
rho \sigma}$
 vanish, the spacetime is  homogenized and isotropized. We use the
decomposition of the Weyl tensor: $E_{\rho \sigma}=~~C_{\rho\mu \sigma
\nu}n^\mu
n^\nu ,  B_{\rho \sigma}=~^\ast C_{\rho\mu \sigma \nu}n^\mu n^\nu$,
where $~^\ast
C_{\mu \nu \rho \sigma}\equiv {1 \over 2}C^{\alpha \beta}_{~~\rho \
sigma}
\varepsilon_{\alpha\beta\mu \nu}$ and $n^\mu$ is a timelike vector
orthogonal to
the hypersurface $\Sigma$.  In  analogy to
 electromagnetism, the 3-dimensional variables $E_{\rho \sigma}$ and
$B_{\rho \sigma}$ are called
an electric and a magnetic parts of the Weyl tensor, and we can
reconstruct the
Weyl tensor completely from this pair
of tensors.  We introduce
\be
{\cal H}(t,x) =E_{\mu \nu}E^{\mu \nu}+B_{\mu \nu}B^{\mu \nu},
\en
as a sort of gravitational ``super-energy" (in fact the purely
timelike component of the
Bel-Robinson tensor\cite{belrob}). If  ${\cal H}(t,x) \rightarrow 0$,
then  $C_{\mu \nu \rho
\sigma} \rightarrow 0$ unless the hypersurfaces become null. We find
that ${\cal
H}(t,x)$ almost vanishes
 within one Hubble expansion time,
 and that this indicates
that our spacetime does indeed approach a homogeneous one ({\it i.e.}
de Sitter spacetime).

In our simulation,  $l$ and ${\cal C}_{\rm max}(0)$ range between
$0.080 l_H \leq l \leq 2.5 l_H$ and $0.020\leq{\cal C}_{\rm max}(0)\
leq 80.0$.
{}From the
present results,  we conclude that for any large Riemann
invariant and/or small width inhomogeneity on the initial
hypersurface, the
nonlinearity of the gravity has little effect and the spacetime always
evolves into a
 de Sitter spacetime.

 Jensen and Stein-Schabes\cite{jensenstein} proved an extension of
Wald's
theorem, showing that inflation invariably eliminates all
inhomogeneities consistent with a
non-positive scalar 3-curvature.  Such inhomogeneities are hardly
generic, however.
 In our models, we find that
 $~^{(3)}R$ is initially negative, but it eventually becomes
positive along the  propagations of the distortions and finally
$~^{(3)}R \rightarrow 0$ when the
spacetime is homogenized (see Figure 2). This convinces us that the
behavior we have found
is not the case  envisioned by Jensen
and Stein-Schabes.

Next, we  simulated  collisions of  two gravitational
pulse waves.  The motivation for such a simulation is the existence of
 exact solutions for colliding plane waves in which  scalar curvature
singularities occur. Szekeres and
Khan-Penrose\cite{colwave} were the first to find such solutions.
Szekeres
treats the collision of two thick or sandwich plane waves, while
Khan-Penrose
treat  two $\delta$ function-like shock waves.  These solutions show
the
nonlinearity of the gravity.  Thus we interest in such a phenomena in
more
general and realistic situations.

It may be interesting to see whether  singularities  also occur when
gravitational waves
collide in the presence of a  cosmological constant.  To investigate
this possibility, we
construct  initial data representing two nearby pulse waves.  Since we
impose  periodic boundary conditions, there is no difference between
the
present case and the case with a single  pulse waves in one  half of
our
numerical region. But in the present setting the collision does not
occur on
the boundary, so it may be better to see the interaction of two waves.
 In Figure 3, a
typical example of the time evolution of the ``curvature" is shown.
We see that the waves start to propagate in opposite directions ($+x$
and $-x$) and
then collide with each other.  When the waves collide, the
``curvature" is just
superposed as in Fig.1, but more clearly.  After the wave-crossing,
each wave
propagates away.
(This behavior is reminiscent of
 solitonic waves.)  Consequently, the spacetime is again homogenized
by
the expansion of the universe within one Hubble expansion time $\
tau_H$.
 For a wide range
of initial widths and ``curvatures" ($0.080
l_H \leq l \leq 0.10 l_H$, $40.0 \leq{\cal C}_{\rm max}(0)\leq
125.0$ and the periodic distance $d$ is  $0.20l_H \leq d \leq
0.50l_H$),  all
inhomogeneities  decay below
 1 \% of their initial ``curvatures" within one Hubble expansion time.

That these spacetime become  homogeneous even if we include
colliding  waves is consistent with the calculations by  Centrella and
Matzner\cite{cenmatz}. They examined the collision of gravitational
shock waves
in an expanding Kasner background both analytically and numerically,
and
concluded that such a collision leads to no  singularity.  The
expansion due to the
cosmological constant in our simulation is exponential --- much faster
than Kasner
(power-law) expansion. Thus, our result would be expected from their
result, although here
we  investigated pulse wave collisions systematically for a wide range
of initial parameters
and extracted the  generic behavior of gravitational waves in
expanding universe.

We also computed the evolution of  [case 2] initial data,
$\hat{\gamma}_{ij} = \delta_{ij}$ and ${\mbox{\rm diag }}(\
hat{K}^{TT}_{ij}) = ( 0, a~{\rm
e}^{-(x/x_0)^2}, -a~{\rm e}^{-(x/x_0)^2}) $,
 and found that all inhomogeneities present in them also decay within
one Hubble expansion
time.  We also tried similar calculations with a different shape for
the initial
pulse waves,
 {\it i.e.}, assuming a  $[1+ a \cos^2 (x/x_0)]$-form for the
$\hat{\gamma}_{yy}$ components, and found the same results for
time evolution.
We  conclude from this that our result is generic: all initial
inhomogeneities decay
within one Hubble expansion time and disappear.

The present  result suggests no  additional condition of the kind
needed to make
 cosmic no hair conjecture a theorem, but it does give a wide class of
examples in
which gravitational inhomogeneities succumb to cosmological constant-
driven expansion.
We could not find any characteristic scale length.   All inhomogeneous
spacetimes expand
to de Sitter universe.    In this sence, our results are different
from  those of
Goldwirth and Piran\cite{GP}, who simulated effects of
 spherically symmetric  inhomogeneities due to
a realistic inflaton  field.  We do not yet know  whether our results
reflect the  generic behavior
of inhomogeneities driven by gravitational waves, or whether they are
largely due
to our {\it ansatz}  of plane symmetry.  In order to answer these
questions,
further simulations including a inflaton field, or more gravitational
degrees of freedom---for example, those present in cylindrical or
axially
symmetrical, or even more general spacetimes----are required.
Such simulations are underway\cite{prepare}.

We would like to acknowledge Dr. L.Gunnarsen for useful discussions
and critical reading
of our manuscript. We also wish to thank K.Nakao and T.Nakamura for
helpful discussions.  This work was supported partially by the Grant-
in-Aid
for Scientific Research Fund of the Ministry of Education, Science and
Culture Nos. (04234211 and 04640312), by a Waseda University Grant for
Special Research Projects and by The Sumitomo Foundation.   This work
was
one of the specific studies at the Center for Informatics, Waseda
University.\par

\newpage

\begin{center}
{\bf Figure Captions}
\end{center}
\vskip 1.0cm
\baselineskip .5in

Figure 1:  \parbox[t]{14.5cm} {
\baselineskip .4in
A typical example of the time evolutions of propagating plane waves.
The ``curvature" ${\cal C}(t,x)$ (see  eq. (\ref{curvature})) is shown
for the
initial pulse shape of $l=0.33l_H$ and ${\cal C}_{\rm max}(t=0)=1.25$
($a=-0.10,
x_0=0.20$ in  eq. (\ref{try1})).
We see the initial pulse propagates both in the $+ x$ and $- x$
directions, but is
smoothed out by the expansion of the universe within one Hubble
expansion time. }

\vskip 0.6cm

Figure 2:  \parbox[t]{14.5cm} {
\baselineskip .4in
The evolution of $~^{(3)}R$ at $x=0$ (center) for the same initial
data as in Fig.1.
This shows also the  maximum deviations from zero at each time.
We see that $~^{(3)}R$ does not remain negative for all time, so that
our simulations
are not governed by the case proven by Jensen and Stein-Schabes.
}
\vskip 0.6cm

Figure 3: \parbox[t]{14.5cm} {
\baselineskip .4in
The time evolution  of the ``curvature"  ${\cal C}(t,x)$ resulting
from
 waves that are located closely. Two waves are the same form,
$l=0.10l_H$ and ${\cal C}_{\rm max}(t=0)=51.0$, and the periodic
distance
is $0.30l_H$. We see them collide  and in the collision region,
the ``curvature" seems to be superposed, but finally the spacetime is
homogenized by the expansion of the universe.}

\end{document}